\documentclass[12pt]{article}

\usepackage{graphicx}
\usepackage{amsmath}
\usepackage{amssymb}
\usepackage{amsthm}
\numberwithin{equation}{section}

\newcommand{\Tr}{\mathrm {Tr\,}}
\newcommand{\Ch}{\mathrm {Ch}}
\newcommand{\sh}{\mathrm {sh}}
\newcommand{\F}{{\mathcal F}}
\newcommand{\C}{{\mathbb C}}
\newcommand{\R}{{\mathbb R}}

\newcommand{\al}{\alpha}
\newcommand{\be}{\beta}
\newcommand{\m}{\mu}

\newcommand{\ld}{\lambda}

\newcommand{\unitbox}
{\setlength{\unitlength}{0.5pt}
\begin{picture}(10,10)
\put(0,10){\line(1,0){10}}
\put(0,0){\line(1,0){10}}
\put(0,0){\line(0,1){10}}
\put(10,0){\line(0,1){10}}
\end{picture}}

\begin{document}

\baselineskip=16pt plus 0.2pt minus 0.1pt

\begin{titlepage}
\title{
\vspace{1cm}
{\Large\bf Instanton Calculus and Loop Operator\\
in Supersymmetric Gauge Theory}
}
\vskip20mm
\author{
{\sc Hiroaki Kanno}\thanks
{{\tt kanno@math.nagoya-u.ac.jp}}\;
\quad and \quad
{\sc Sanefumi Moriyama}\thanks
{{\tt moriyama@math.nagoya-u.ac.jp}}\\[15pt]
{\it Graduate School of Mathematics,} \\
{\it Nagoya University,} \\
{\it Nagoya 464-8602, Japan}
}
\date{\normalsize December, 2007}
\maketitle
\thispagestyle{empty}
\begin{abstract}
\normalsize
We compute one-point function of the glueball loop operator in the
maximally confining phase of supersymmetric gauge theory using instanton
calculus.
In the maximally confining phase the residual symmetry is the diagonal
$U(1)$ subgroup and the localization formula implies that the chiral
correlation functions are the sum of the contributions from each fixed
point labeled by the Young diagram.
The summation can be performed exactly by operator formalism of free
fermions, which also featured in the equivariant Gromov-Witten theory of
${\bf P}^1$.
By taking the Laplace transformation of the glueball loop operator, we
find an exact agreement with the previous results for the generating
function (resolvent) of the glueball one-point functions.
\end{abstract}

\end{titlepage}

\section{Introduction}
Symmetry helps us to understand physics, in particular, the
non-perturbative dynamics.
Guided by holomorphy coming from ${\cal N}=2$ supersymmetry and the
asymptotic behavior in semiclassical region, Seiberg and Witten wrote
down the effective prepotential of ${\cal N}=2$ supersymmetric gauge
theory \cite{SW}.
However, it is still desirable to derive their result following the
standard process of quantum field theory - path integral.
After a lot of works on direct integration over the moduli space of
instantons, Nekrasov \cite{N} applied the localization theorem for 
toric actions to compute the equivariant integral over the instanton
moduli space.
Subsequently, it was noted by \cite{FMPT} that a similar approach also
applies to the instanton calculus of ${\cal N}=1$ supersymmetric gauge
theory where ${\cal N}=2$ theory is perturbed by a superpotential
$W(\Phi)$.
Here the integration is localized to the summation over the (isolated)
fixed points labeled by the Young diagrams, which are in some sense
regarded as saddle points.
One might expect the instanton calculus on the saddle points has more
applications than kinematical constraints.

The computation of the holomorphic quantities in ${\cal N}=1$ theory
such as effective superpotential is related to matrix model in
\cite{DV}.
This proposal is derived by relating them to the B-model topological
string amplitudes on certain non-compact Calabi-Yau manifold.
The relation is a mirror to a gauge/string duality due to the geometric
transition on the A-model side \cite{GV}.
Thus the reproduction of the holomorphic quantities in ${\cal N}=1$
theory from the microscopic instanton calculus can be regarded as a
check of the gauge/string correspondence involving the matrix model.
For a recent progress in microscopic approach to ${\cal N}=1$
supersymmetric gauge theory, see \cite{Fer1,FKW,Fer2}.

Though the localization formulas are given explicitly for both the
partition function and the chiral correlation functions, in general it
is not straightforward to perform the summation over the Young diagrams
exactly.
In the maximally confining phase of $U(N)$ supersymmetric gauge
theory\footnote{We reduce the computation to the instanton calculus of
$U(1)$ gauge theory. However, the existence of underlying $U(N)$ theory
should be assumed, since we will consider the expansion in the second
Chern number whose meaning is lost in a genuine $U(1)$ theory.},
where $SU(N)$ is confined and the residual symmetry at low energy is the
diagonal $U(1)\subset U(N)$, the task of summation gets considerably
tractable.
In \cite{FKMO}, we adopted the standard correspondence between Young
diagrams and neutral states in the fermion Fock space and calculated
explicitly the chiral one-point functions $\langle\Tr\varphi^{J}\rangle$
in the maximally confining phase.
We found that the results can be summarized compactly in terms of the
loop operator:
\begin{align}
\langle\Tr e^{u\varphi}\rangle
=I_0\left(4\sqrt{q}~\frac{\sinh(u\hbar/2)}{\hbar} \right)
\to I_0(2\sqrt{q}u)~,\qquad(\hbar\to 0)~,\label{loop}
\end{align}
with $I_n(x)$ being the modified Bessel function.
The vacuum expectation value of the loop operator has two parameters
$q=\Lambda^{2N}$ and $\hbar$.
The series expansion in $q$ is the instanton expansion of gauge theory,
while that in $\hbar$ is interpreted as the genus expansion of the
corresponding string theory.
In the gauge/string theory correspondence, the partition function
and the correlation functions allow the double perturbative expansion in
the instanton number of gauge theory and the genus of the string theory.
Usually we can sum up the expansion only in one of the two expansion
parameters, but have to compute the functions order by order in the
other parameter.
It is very amusing that the above result \eqref{loop} gives an example
where we can perform the summation of the double expansion completely
in a closed form.

The chiral one-point functions $\langle\Tr\varphi^{J}\rangle$ in 
${\cal N}=1$ theory do not detect the superpotential $W(\Phi)$ which is
used to deform the original ${\cal N}=2$ theory \cite{FMPT,FKMO}.
In this paper we would like to proceed to the computation of the chiral
glueball one-point functions
$\langle\Tr\lambda^\alpha\lambda_\alpha\varphi^{J}\rangle$, which depend
on the details of $W(\Phi)$.
The chiral glueball one-point functions were obtained previously by
several methods.
For example, one can relate the holomorphic ($F$-term) quantities in
supersymmetric gauge theory to amplitudes of topological string theory
\cite{BCOV} and then compute with matrix model using open/closed string
duality \cite{DV}.
We can also relate the correlation functions to the generalized Konishi
anomaly and perform a purely field theoretical computation \cite{CDSW}.
The resulting glueball resolvent is given by
\begin{align}
R(z):=\langle\Tr\lambda^\alpha\lambda_\alpha\frac{1}{z-\varphi}\rangle
=-W'(z)+\sqrt{W'(z)^2-4f(z)}~,\label{resolvent}
\end{align}
with an unknown polynomial $f(z)$ which is determined so that the
asymptotic behavior of $R(z)$ in $z\to\infty$ is 
$\langle\Tr\lambda^\alpha\lambda_\alpha\rangle\cdot z^{-1}+O(z^{-2})$.
In the maximally confining phase, the polynomial $f(z)$ is further tuned
so that the square root is factorized into
\begin{align}
\sqrt{W'(z)^2-4f(z)}=H(z)\sqrt{(z-a)^2-4q}~.
\label{factorize}
\end{align}
In matrix model the maximally confining phase is described by a one-cut
solution, while on the moduli space of Seiberg-Witten theory it
corresponds to the degenerating loci where the maximal number of
mutually non-intersecting cycles of the Seiberg-Witten curve collapse.
In this paper we would like to add one more method to derive the
glueball resolvent $R(z)$.
We find our instanton calculus gives an exact agreement to the result
\eqref{resolvent} with the factorization \eqref{factorize}.
The reproduction of previously known results should serve as a
non-trivial consistency check of the instanton calculus.
Besides, along the way of computation we also find an explicit
expression of the polynomial $f(z)$ that causes the complete
factorization \eqref{factorize}.
It is interesting that our $U(1)$ instanton calculus automatically gives
a closed form for the glueball resolvent in the maximally confining
phase.

The plan of the present paper is as follows.
We will shortly review some necessary ingredients of the instanton
calculus in section 2.
Before proceeding to our computation of the chiral glueball one-point
functions in section 4, we will first compute two-point functions of the
loop operators as a preparation in section 3.
Finally in section 5 we comment on the higher genus correction to our
computation.

\section{Review of instanton calculus}

Chiral operators ${\mathcal O}$ in supersymmetric field theories are, by
definition, those annihilated by the fermionic charges
$\overline{Q}_{\dot\alpha}$ of one chirality.
Two chiral operators are defined to be equivalent, if the difference
is $\overline{Q}_{\dot\alpha}$-exact.
The set of chiral operators is closed under the multiplication and form
a ring, which we call chiral ring.
{}From the supersymmetry algebra in four dimensions,
$\{Q_\alpha,\overline{Q}_{\dot\alpha}\}
=\sigma_{\alpha\dot\alpha}^\m P_\m$,
we can see that correlation functions of chiral operators are
``topological'' in the sense that they are independent of the positions
of operators.
Especially, topological one-point functions characterize the phase
structure of vacua.
In the four dimensional ${\mathcal N}=1$ supersymmetric gauge theory
with a single adjoint matter $\Phi$, we have a vector multiplet
$(A_\mu,\ld_\alpha)$ and a chiral multiplet
$\Phi=(\varphi,\psi_\alpha)$, which are all in the adjoint
representation.
One can show that the generators of chiral ring are of the form
$\Tr\varphi^J$, $\Tr\ld_\alpha\varphi^J$ and
$\Tr\ld^\alpha\ld_\alpha\varphi^J$ \cite{CDSW}.
The correlation functions of the chiral operators ${\mathcal O}$ are
defined by
\begin{align}
\langle{\mathcal O}\rangle_{{\mathcal N}=1}
:=\frac{1}{V{\mathcal Z}_{{\mathcal N}=1}}\int_{{\mathcal M}}
\left\{\int_{\C^2}{\mathcal O}\right\}\exp(-S_{{\mathcal N}=1})~,
\label{correldef}
\end{align}
where the action for ${\mathcal N}=1$ supersymmetric gauge theory
$S_{{\mathcal N}=1}$ is that of ${\mathcal N}=2$ theory
$S_{{\mathcal N}=2}$ perturbed by a superpotential $W(\Phi)$:
\begin{align}
S_{{\mathcal N}=1}
:=S_{{\mathcal N}=2}+\int dx^4d\theta^2W(\Phi)~.
\end{align}
The correlator $\langle{\mathcal O}\rangle_{{\mathcal N}=1}$ is
normalized by the volume $V$ of the non-commutative $\C^2$ and the
partition function ${\mathcal Z}_{{\mathcal N}=1}
:=\int_{{\mathcal M}}\exp(-S_{{\mathcal N}=1})$.
The integral is over the moduli space ${\mathcal M}$ of $U(N)$
instantons described, for example, by the ADHM construction.
The moduli space has the decomposition 
${\mathcal M}:=\sqcup_k{\mathcal M}_{N,k}$, that is, the correlation
function is a sum over the contributions from each moduli space
${\mathcal M}_{N,k}$ with a specific instanton number $k$.
The correlation function $\langle{\mathcal O}\rangle_{{\mathcal N}=1}$
for ${\mathcal O}=\Tr\varphi^J$ or
$\Tr\lambda^{\alpha}\lambda_{\alpha}\varphi^J$ is of our prime interest
in this paper.
We shall omit the subscripts ${\mathcal N}=1$ or ${\mathcal N}=2$ of the
correlators hereafter as long as it is clear from the context.

The strategy to calculate these correlation functions is to first
replace all the operators by their equivariant extensions \cite{FMPT}:
\begin{align}
\Tr\varphi^J&\mapsto\alpha_{(2,2)}\wedge\Tr{\mathcal F}^J~,\\
\Tr\lambda^{\alpha}\lambda_{\alpha}\varphi^{J}&\mapsto
-\frac{1}{(J+2)(J+1)}\alpha_{(0,2)}\wedge\Tr{\mathcal F}^{J+2}~,\\
d^2\theta\,W(\Phi)&\mapsto\alpha_{(2,0)}\wedge\Tr W({\mathcal F})~,
\end{align}
with $\alpha$ being the equivariantly closed forms
\begin{align}
&\alpha_{(0,0)}:=1~,\\
&\alpha_{(2,0)}:=dz^1\wedge dz^2+i\hbar z^1z^2~,\\
&\alpha_{(0,2)}:=d\overline{z}^1\wedge d\overline{z}^2
-i\hbar\overline{z}^1\overline{z}^2~,\\
&\alpha_{(2,2)}:=\alpha_{(2,0)}\wedge\alpha_{(0,2)}~.
\end{align}
The curvature ${\mathcal F}$ of the universal bundle over 
$\C^2\times{\mathcal M}_{N,k}$ is expanded according to the direct
product structure of the base space as follows\footnote{The matrix $U$
appears in the ADHM construction as the zero modes of the Dirac
operator.}:
\begin{align}
{\mathcal F}&=:F+\Psi+\varphi\nonumber\\
&=F_{\mu\nu}dx^{\mu}dx^{\nu}
+\{\lambda_mdz^m+\psi_{\overline{m}}d\overline{z}^{\overline{m}}\}
+\{(d_{{\mathcal M}}U^{\dagger})(d_{{\mathcal M}}U)
-U^{\dagger}{\mathcal L}_{\xi}U\}~.
\end{align}
If we untwist the theory\footnote{Note that we consider the case
$\C^2\simeq\R^4$ in this paper. In general the topological twist of
${\mathcal N}=1$ theory is possible on the K\"ahler manifold.}, the
components $F_{\mu\nu}$ and $\lambda_m$ are identified with the field
strength and the gaugino, which gives a vector multiplet in 
${\mathcal N}=1$ theory.
The remaining pair $\psi_{\overline{m}}$ and $\varphi$ gives a chiral
multiplet.
Then due to the localization theorem, our computation reduces to picking
up the value at the fixed points.
Since the fixed points are classified by the Young diagrams, the result
is given in terms of summation over the Young diagrams.
Especially in the maximally confining phase, the correlation functions
\eqref{correldef} simplifies into
\begin{align}
\langle{\mathcal O}\rangle=\frac{1}{{\mathcal Z}}
\sum_{k=0}^\infty\biggl[\frac{q}{\hbar^2}\biggr]^k\sum_{|Y|=k}
\frac{{\mathcal O}_Y}{\prod_{\unitbox\in Y}(h(\unitbox))^2}~,
\end{align}
where the partition function is given by
\begin{align}
{\mathcal Z}=\sum_{k=0}^\infty\biggl[\frac{q}{\hbar^2}\biggr]^k
\sum_{|Y|=k}\frac{1}{\prod_{\unitbox\in Y}(h(\unitbox))^2}~.
\label{ZU1}
\end{align}
We have introduced the parameter $q$ of the instanton expansion.
$\hbar$ is the equivariant parameter associated with the $U(1)$ action
$T_{\hbar}:(z_1,z_2)\to(e^{i\hbar}z_1,e^{-i\hbar}z_2)$ on $\C^2$. 
Note that the partition function ${\mathcal Z}$ depends on these
parameters only through the combination $q/\hbar^2$.
We denote by $|Y|$ the total number of boxes of a Young diagram $Y$.
The hook length at a box $\unitbox\in Y$ is denoted by $h(\unitbox)$
and the weight $\prod_{\unitbox\in Y}(h(\unitbox))^{-2}$ is called the
Plancherel measure. 
At each fixed point $Y$ we can estimate the chiral operator 
${\mathcal O}$ to obtain ${\mathcal O}_Y$.
Thus ${\mathcal O}_Y$ is a function on the space of Young diagrams and
$\langle{\mathcal O}\rangle$ is nothing but the integration of
${\mathcal O}_Y$ with respect to the Plancherel measure\footnote{The
Plancherel measure can be regarded as a discretization of the
Vandermonde measure. This interpretation suggests a natural connection
to the matrix model.}.

For the loop operator ${\mathcal O}=\Tr e^{t\varphi}$, which is a
generating function of $\Tr\varphi^J$, the function ${\mathcal O}_Y$ is
given by
\begin{align}
\Ch_Y(\alpha)=e^{ta}\Bigl(1+\sh^2(\al)
\sum_{\unitbox\in Y}e^{\alpha(c(\unitbox))}\Bigr)~,
\end{align}
where $a=\langle\Tr\varphi\rangle$, $\al=t\hbar$ and
$\sh(\al):=e^{\al/2}-e^{-\al/2}$ \cite{FKMO}.
For a box $\unitbox$ at the $m$-th row and the $n$-th column of the
Young diagram, we define the content by $c(\unitbox):=n-m$.
In \cite{FMPT} the glueball operator 
${\mathcal O}=\Tr\lambda^\alpha\lambda_\alpha\varphi^J$ is related to
the connected two-point function by
\begin{align}
\langle\Tr\lambda^\alpha\lambda_\alpha\varphi^J\rangle_{{\mathcal N}=1}
=-\frac{2}{(J+2)(J+1)\hbar^2}
\langle\Tr W(\varphi)\,\Tr\varphi^{J+2}\rangle_{\rm conn}~.
\label{gluinoJ}
\end{align}
The relation \eqref{gluinoJ} may be derived as follows\footnote{Two 
proofs of \eqref{gluinoJ} have been presented recently in
\cite{Fer1,Fer2}.}:
The definition \eqref{correldef} implies that the ${\mathcal N}=1$
correlators are related to the ${\mathcal N}=2$ correlators by
\begin{align}
\langle{\mathcal O}\rangle_{{\cal N}=1}
=\langle{\mathcal O}\rangle_{{\cal N}=2}
+\langle{\mathcal O}\int_{{\mathbb C}^2}\alpha_{(2,0)}\wedge\Tr W(\F)
\rangle_{{\cal N}=2}+\cdots~.
\end{align}
Hence we have
\begin{align}
\langle\int_{{\mathbb C}^2}\alpha_{(0,2)}\wedge\Tr\F^{J+2}
\rangle_{{\cal N}=1}
&=\hbar\bar{z}^1\bar{z}^2\langle\Tr\varphi^{J+2}\rangle
+\hbar^2z^1z^2\bar{z}^1\bar{z}^2
\langle\Tr\varphi^{J+2}\,\Tr W(\varphi)\rangle+\cdots~,\\
V\langle 1\rangle_{{\cal N}=1}
&=\hbar^2z^1z^2\bar{z}^1\bar{z}^2\Bigl(\langle 1\rangle
+\hbar z^1z^2\langle\Tr W(\varphi)\rangle+\cdots\Bigr)~.
\end{align}
Note that in the expansion the first term vanishes after setting
$\bar{z}^1,\bar{z}^2\to 0$, while the higher-order terms vanish because
of $z^1,z^2\to 0$.
Summing up all the terms in \eqref{gluinoJ} into the exponential
function, we find that the glueball loop operator is given as
\begin{align}
\langle\Tr\lambda^\alpha\lambda_\alpha
e^{u\varphi}\rangle_{{\mathcal N}=1}
=-\frac{2}{\hbar^2}
\langle\Tr W(\varphi)\,
\Tr\frac{e^{u\varphi}-1-u\varphi}{u^2}\rangle_{\rm conn}~.
\label{gluino1+u}
\end{align}

The more conventional quantity is the glueball resolvent, which is the
Laplace transformation of the glueball loop operator
\begin{align}
R(z):=\langle\Tr\lambda^\alpha\lambda_\alpha\frac{1}{z-\varphi}\rangle
=\int_0^\infty du~e^{-zu}
\langle\Tr\lambda^\alpha\lambda_\alpha e^{u\varphi}\rangle~.
\end{align}
The explicit form of the glueball resolvent is known to be
\begin{align}
R(z)=-W'(z)+\sqrt{W'(z)^2-4f(z)}~,
\label{known}
\end{align}
{}from the relation to the matrix model \cite{DV} or the generalized
Konishi anomaly \cite{CDSW}.
For the superpotential $W(z)$ of degree $n+1$, we need to include a
polynomial $f(z)$ of degree $n-1$.
Since we are considering the maximally confining phase, we have to
choose the coefficients of $f(z)$ so that $\sqrt{W'(z)^2-4f(z)}$ is
factorized into a product of a polynomial and the square root of a
quadratic function.
Therefore, the resolvent $R(z)$ in the maximally confining phase is
characterized by the following conditions:
\begin{itemize}
\item
The resolvent $R(z)$ is a linear combination of $1$ and the square root
of a quadratic function with the coefficients being polynomials of $z$.
\item
The behavior of $z\to\infty$ is
$\langle\Tr\lambda^\alpha\lambda_\alpha\rangle\cdot z^{-1}+O(z^{-2})$.
\item
The coefficient of $1$ is $-W'(z)$.
\end{itemize}
We would like to reproduce all these features from the following
instanton calculus.

\section{Scalar two-point functions}

In this section we compute the two-point correlation function of the
loop operator:
\begin{align}
\langle\Tr e^{t\varphi}\,\Tr e^{u\varphi}\rangle
=\frac{1}{{\mathcal Z}}
\sum_{k=0}^\infty\biggl[\frac{q}{\hbar^2}\biggr]^k
\sum_{|Y|=k}\frac{\Ch_Y(\al)\Ch_Y(\be)}
{\prod_{\unitbox\in Y}(h(\unitbox))^2}~,
\end{align}
where
\begin{align}
\Ch_Y(\al)=1+\sh^2(\al)\sum_{\unitbox\in Y}e^{\al(c(\unitbox))}~,
\end{align}
with $\al=t\hbar$ and $\be=u\hbar$.
(We set $a=0$ for simplicity in this section, but we can easily
reproduce the contribution by multiplying the final result by the
factors $e^{ta}$ and $e^{ua}$.)
Taking care of the constant term separately, we find
\begin{align}
\langle\Tr e^{t\varphi}\,\Tr e^{u\varphi}\rangle={\mathcal W}
+\langle\Tr e^{t\varphi}\rangle+\langle\Tr e^{u\varphi}\rangle-1~,
\end{align}
where
\begin{align}
{\mathcal W}:=\frac{1}{{\mathcal Z}}
\sum_{k=0}^\infty\biggl[\frac{q}{\hbar^2}\biggr]^k
\sh^2(\al)\sh^2(\be)\sum_{|Y|=k}
\frac{\sum_{\unitbox\in Y}e^{\al(c(\unitbox))}
\sum_{\unitbox\in Y}e^{\be(c(\unitbox))}}
{\prod_{\unitbox\in Y}(h(\unitbox))^2}~.
\label{separate}
\end{align}

\subsection{Operator formalism}

It is well-known that there is a correspondence between the Young
diagrams and the fermion Fock states with neutral charge.
By making use of the correspondence, we can compute the summations over
the set of Young diagrams in operator formalism.
Let us introduce a pair of charged (NS) free fermions
\begin{align}
\psi(z)=\sum_{r\in{\mathbb Z}+\frac{1}{2}}\psi_rz^{-r-\frac{1}{2}}~,
\quad
\psi^*(z)=\sum_{s\in{\mathbb Z}+\frac{1}{2}}\psi^*_sz^{-s-\frac{1}{2}}~,
\end{align}
with the anti-commutation relation
\begin{align}
\{\psi_r,\psi_s^*\}=\delta_{r+s,0}~,\quad 
r,s\in{\mathbb Z}+\frac{1}{2}~.
\label{ACR}
\end{align}
We define the Fock vacuum $|0\rangle$ by
\begin{align}
\psi_r|0\rangle=\psi^*_s|0\rangle=0~,\quad r,s>0~.
\end{align}
Recall that the Young diagram is specified by a partition $\ld=(\ld_i)$,
where $\ld_i$ is the length of the $i$-th row.
Then the corresponding state is given by
\begin{align}
|\ld\rangle=\prod_{i=1}^\infty
\psi_{i-\ld_i-\frac{1}{2}}|\!|0\rangle\!\rangle~,
\end{align}
with $\psi_s^*|\!|0\rangle\!\rangle=0,~\forall s$.
One can show that
\begin{align}
J_{-1}^k|0\rangle=\sum_{|\lambda|=k}
\frac{k!}{\prod_{\unitbox\in\lambda}h(\unitbox)}|\lambda\rangle~,
\label{J-1power}
\end{align}
where $J_{-1}$ is the constant mode in the standard $U(1)$ current
$J(z):=:\psi(z)\psi^*(z):$.
{}From \eqref{J-1power} we can confirm the following famous relation
\begin{align}
\sum_{|Y|=k}\prod_{\unitbox\in Y}h(\unitbox)^{-2}=\frac{1}{k!}~,
\label{norm}
\end{align}
which is used for computing the summation over the Young diagrams
appearing in the partition function ${\mathcal Z}$.
We find \eqref{ZU1} has a simple form:
\begin{align}
{\mathcal Z}=\exp\left(\frac{q}{\hbar^2}\right)~.
\end{align}
This is the most fundamental example of the computation in operator
formalism of the summation with the Plancherel measure \cite{NO}.

To compute the correlation functions, it is convenient to introduce the
operator \cite{OP1,OP2}
\begin{align}
{\mathcal E}^{(\al)}(z):
=:\psi(e^{\frac{\al}{2}}z)\psi^*(e^{-\frac{\al}{2}}z):
=\sum_{n\in{\mathbb Z}}{\mathcal E}_n^{(\al)}z^{-n-1}~,
\end{align}
which is a ``point splitting'' deformation of 
$J(z)={\mathcal E}^{(0)}(z)$.
The modes of ${\mathcal E}^{(\al)}(z)$
\begin{align}
{\mathcal E}_n^{(\al)}:=\sum_{r\in{\mathbb Z}+\frac{1}{2}}
e^{\al\left(r-\frac{n}{2}\right)}:\psi_{n-r}\psi_r^*:~,
\label{OPop}
\end{align}
satisfy the commutation relation\footnote{The infinite-dimensional
Lie algebra with the commutation relation \eqref{Wcom} appeared
first in \cite{FFZ}, where it was called area-preserving torus
diffeomorphism algebra. The representation theory 
was initiated in \cite{KR} and the algebra was 
identified with $W_{1+\infty}$ algebra. Recently the same algebra
is used to reveal a connection of the counting of plane partitions and 
the Toda hierarchy \cite{NT}.},
\begin{align}
[{\mathcal E}_n^{(\al)},{\mathcal E}_m^{(\be)}]
={\sh}(n\be-m\al)~{\mathcal E}_{n+m}^{(\al+\be)}
+\delta_{n+m,0}\frac{\sh\,n(\al+\be)}{\sh\,(\al+\be)}~.
\label{Wcom}
\end{align}
We will use the following important relation in the computation of
correlation functions:
\begin{align}
{\mathcal E}_0^{(\al)}|\lambda\rangle
=\sum_{i=1}^\infty\left(e^{\al\left(\lambda_i-i+\frac{1}{2}\right)}
-e^{\al\left(-i+\frac{1}{2}\right)}\right)|\lambda\rangle
={\sh(\al)}\sum_{\unitbox\in\lambda}
\exp\left(\al c(\unitbox)\right)|\lambda\rangle~.
\end{align}
The second term comes from ${\cal E}_0^{(\al)}|\!|0\rangle\!\rangle
=-({\mathrm{sh}}(\al))^{-1}|\!|0\rangle\!\rangle$, which can be
calculated directly from the definition of $|\!|0\rangle\!\rangle$
or from the consistency ${\cal E}_0^{(\al)}|0\rangle=0$.

\subsection{Computation}

After the preparation of the operator formalism, let us proceed to the
explicit calculation of ${\mathcal W}$ defined by \eqref{separate}.
First of all, note that the formulas in the operator formalism imply
\begin{align}
\langle 0|J_1^k{\cal E}_0^{(\al)}{\cal E}_0^{(\be)}J_{-1}^k|0\rangle
=(k!)^2\sh(\al)\sh(\be)\sum_{|Y|=k}
\frac{\sum_{\unitbox\in Y}e^{\al(c(\unitbox))}\sum_{\unitbox\in
Y}e^{\be(c(\unitbox))}}
{\prod_{\unitbox\in Y}(h(\unitbox))^2}~.
\end{align}
Thus the problem of evaluation of ${\mathcal W}$ reduces to the
computation of $\langle 0|J_1^k{\cal E}_0^{(\al)}
{\cal E}_0^{(\be)}J_{-1}^k|0\rangle$.
With the help of the commutation relation \eqref{Wcom} with $J_{1}
={\cal E}_{1}^{(0)}$, we obtain
\begin{align}
\langle 0|J_1^k{\cal E}_0^{(\al)}{\cal E}_0^{(\be)}J_{-1}^k|0\rangle
=\sum_{l=0}^k\sum_{m=0}^{k-l}
\frac{(k!)^2\sh^l(\al)\sh^m(\be)}{l!m!(k-l-m)!(l+m)!}
\langle 0|{\cal E}_l^{(\al)}{\cal E}_m^{(\be)}J_{-1}^{l+m}|0\rangle~,
\end{align}
where we have used
\begin{align}
J_{1}^{k-l-m}J_{-1}^k|0\rangle
=\frac{k!}{(l+m)!}J_{-1}^{l+m}|0\rangle~.
\end{align}
Bringing $J_{-1}$'s to the most left in $\langle 0|\cdots|0\rangle$ by
iteratively using the commutation relation
\begin{align}
[{\cal E}_l^{(\al)}{\cal E}_m^{(\be)},J_{-1}]
={\cal E}_l^{(\al)}
\bigl(\sh(\be){\cal E}_{m-1}^{(\be)}+\delta_{m,1}\bigr)
+\bigl(\sh(\al){\cal E}_{l-1}^{(\al)}+\delta_{l,1}\bigr)
{\cal E}_m^{(\be)}~,
\label{comm}
\end{align}
we are left with a constant term and terms such as
$\langle 0|{\cal E}_n^{(\al)}{\cal E}_{-n}^{(\be)}|0\rangle$.
The coefficients of these terms can be calculated as follows.
We first place the operator ${\cal E}_l^{(\al)}{\cal E}_m^{(\be)}$ at
the point $(l,m)$ in the two-dimensional plane.
(See Figure 1.)
Using the commutation relation \eqref{comm} once means that we bring the
operator downwards or leftwards by one unit in the plane.
After $(l+m)$-times of the iterative moves, we finally arrive at the
integer point $(n,-n)$.
The coefficient of the
$\langle 0|{\cal E}_n^{(\al)}{\cal E}_{-n}^{(\be)}|0\rangle$ term (or
the constant term) can be interpreted as the combinatorial factor of
moving from $(l,m)$ to $(n,-n)$ (or $(0,0)$ respectively) by these
iterative moves.
\begin{figure}[htbp]
\begin{center}
\scalebox{1.0}[1.0]{\includegraphics{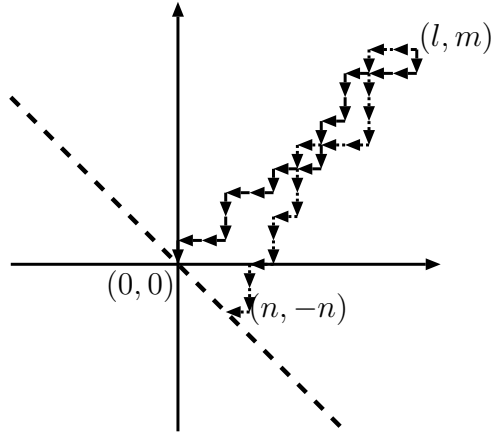}}
\setlength{\unitlength}{1mm}
\begin{picture}(20,20)
\put(-6,48){\makebox(10,10){$(l,m)$}}
\put(-48,15){\makebox(10,10){$(0,0)$}}
\put(-27,12){\makebox(10,10){$(n,-n)$}}
\end{picture}
\end{center}
\caption{Combinatorics of the coefficients.}
\end{figure}
In this way we find
\begin{align}
\frac{\langle 0|{\cal E}_l^{(\al)}{\cal E}_m^{(\be)}
J_{-1}^{l+m}|0\rangle}{(l+m)!}
=\sum_{n=0}^l\frac{[\sh(\al)]^{l-n}[\sh(\be)]^{m+n}}{(l-n)!(m+n)!}
\frac{\sh\,n(\al+\be)}{\sh\,(\al+\be)}
+\frac{[\sh(\al)]^{l-1}[\sh(\be)]^{m-1}}{l!m!}
\Theta_{l>0}\Theta_{m>0}~.
\end{align}
Note that the second term, which comes from the constant term in the
commutation relation \eqref{comm}, contributes only when $l\ne 0$ and
$m\ne 0$.
Plugging back to \eqref{separate}, we call each contribution from the
above two terms as ${\mathcal W}_1$ and ${\mathcal W}_2$, respectively.

Let us concentrate on ${\mathcal W}_2$ first.
\begin{align}
{\mathcal W}_2=\frac{1}{{\mathcal Z}}
\sum_{k=0}^\infty\sum_{l=0}^k\sum_{m=0}^{k-l}
\biggl[\frac{q}{\hbar^2}\biggr]^k
\frac{[\sh(\al)]^{2l}[\sh(\be)]^{2m}}{(l!)^2(m!)^2(k-l-m)!}
\Theta_{l>0}\Theta_{m>0}~.
\end{align}
We can exchange the order of summation by bringing the $k$-summation
into the most right
\begin{align}
\sum_{k=0}^\infty\sum_{l=0}^k\sum_{m=0}^{k-l}\cdots
=\sum_{l=0}^\infty\sum_{k=l}^\infty\sum_{m=0}^{k-l}\cdots
=\sum_{l=0}^\infty\sum_{m=0}^\infty\sum_{k=l+m}^\infty\cdots
\label{exchange}
\end{align}
After the exchange of summation, we find ${\mathcal W}_2$ is computed to
be
\begin{align}
{\mathcal W}_2=\bigl(I_0(A)-1\bigr)\bigl(I_0(B)-1\bigr)~,
\end{align}
with $A=2\sqrt{q}\,\sh(\al)/\hbar,~B=2\sqrt{q}\,\sh(\be)/\hbar$
and $I_n(x)$ is the $n$-th modified Bessel function.

Now let us turn to ${\mathcal W}_1$:
\begin{align}
{\mathcal W}_1=\frac{1}{{\mathcal Z}}
\sum_{k=0}^\infty\sum_{l=0}^k\sum_{m=0}^{k-l}\sum_{n=0}^l
\biggl[\frac{q}{\hbar^2}\biggr]^k
\frac{[\sh(\al)]^{2l-n+1}[\sh(\be)]^{2m+n+1}}
{l!(l-n)!m!(m+n)!(k-l-m)!}
\frac{\sh\,n(\al+\be)}{\sh\,(\al+\be)}~.
\end{align}
Here the exchange of summation goes as
\begin{align}
\sum_{k=0}^\infty\sum_{l=0}^k\sum_{m=0}^{k-l}\sum_{n=0}^l\cdots
=\sum_{l=0}^\infty\sum_{m=0}^\infty\sum_{k=l+m}^\infty
\sum_{n=0}^l\cdots
=\sum_{l=0}^\infty\sum_{n=0}^l\sum_{m=0}^\infty
\sum_{k=l+m}^\infty\cdots
=\sum_{n=0}^\infty\sum_{l=n}^\infty\sum_{m=0}^\infty
\sum_{k=l+m}^\infty\cdots
\end{align}
where we have used \eqref{exchange} in the first equality and then
brought the $n$-summation to the most left.
Performing the $k$-summation, we have
\begin{align}
{\mathcal W}_1=\sum_{n=0}^\infty
\frac{\sh\,n(\al+\be)}{\sh\,(\al+\be)}
\biggl(\sum_{l=n}^\infty\biggl[\frac{q}{\hbar^2}\biggr]^l
\frac{[\sh(\al)]^{2l-n+1}}{l!(l-n)!}\biggr)
\biggl(\sum_{m=0}^\infty\biggl[\frac{q}{\hbar^2}\biggr]^m
\frac{[\sh(\be)]^{2m+n+1}}{m!(m+n)!}\biggr)~,
\end{align}
which implies
\begin{align}
{\mathcal W}_1=\sh(\al)\sh(\be)\sum_{n=0}^\infty
I_n(A)I_n(B)\frac{\sh\,n(\al+\be)}{\sh\,(\al+\be)}~.
\end{align}

Plugging back into \eqref{separate} all our results including the
one-point function $\langle\Tr e^{t\varphi}\rangle=I_0(A)$,
we finally find that the two-point function is given by
\begin{align}
\langle\Tr e^{t\varphi}\,\Tr e^{u\varphi}\rangle
=I_0(A)I_0(B)
+\sh(\al)\sh(\be)
\sum_{n=0}^\infty I_n(A)I_n(B)\frac{\sh\,n(\al+\be)}{\sh\,(\al+\be)}~.
\end{align}
Note that the first term can be interpreted as the disconnected
contribution.
Hence,
\begin{align}
\langle\Tr e^{t\varphi}\,\Tr e^{u\varphi}\rangle_{\rm conn}
&:=\langle\Tr e^{t\varphi}\,\Tr e^{u\varphi}\rangle
-\langle\Tr e^{t\varphi}\rangle\langle\Tr e^{u\varphi}\rangle\nonumber\\
&\:=\sh(\al)\sh(\be)
\sum_{n=0}^\infty I_n(A)I_n(B)\frac{\sh\,n(\al+\be)}{\sh\,(\al+\be)}~.
\label{twopt}
\end{align}

\section{Glueball one-point functions}

Having finished our computation of the scalar two-point function in the
previous section, let us turn to the glueball one-point function.
As \eqref{gluino1+u} implies, all the information we need is encoded in
the scalar two-point function \eqref{twopt} in the limit $\hbar\to 0$:
\begin{align}
\lim_{\hbar\to 0}\frac{1}{\hbar^2}\langle\Tr e^{t\varphi}\,
\Tr e^{u\varphi}\rangle_{\rm conn}
=e^{ta}e^{ua}\,tu\sum_{n=0}^\infty nI_n(2\sqrt{q}t)I_n(2\sqrt{q}u)~,
\label{looploop}
\end{align}
where we have restored the classical value $a=\langle\Tr\varphi\rangle$
by simply multiplying the factors $e^{ta}$ and $e^{ua}$.
Using this result, we find only the exponential term in
\eqref{gluino1+u} gives a non-trivial contribution:
\begin{align}
\langle\Tr\lambda^\alpha\lambda_\alpha e^{u\varphi}\rangle
=\lim_{\hbar\to 0}-\frac{2}{\hbar^2u^2}\langle\Tr W(\varphi)
\,\Tr e^{u\varphi}\rangle_{\rm conn}~.
\label{gluino}
\end{align}
Then, all we have to do is to pick up necessary terms out of
\eqref{looploop} and perform the Laplace transformation to obtain the
glueball resolvent.

\subsection{An example}

Before proceeding to the general superpotential, let us consider
a simple example
\begin{align}
W(z)=\frac{z^4}{4}~,
\end{align}
which allows an explicit computation and is instructive for general
cases.
Since the connected two-point function is given by
\begin{align}
&\lim_{\hbar\to 0}\frac{1}{\hbar^2}
\langle\Tr e^{t\varphi}\,\Tr e^{u\varphi}\rangle_{\rm conn}
=e^{ta}e^{ua}tu\nonumber\\
&\quad\times\Bigl(I_1(2\sqrt{q}t)I_1(2\sqrt{q}u)
+2I_2(2\sqrt{q}t)I_2(2\sqrt{q}u)
+3I_3(2\sqrt{q}t)I_3(2\sqrt{q}u)+\cdots\Bigr)~,
\end{align}
with the modified Bessel functions being
\begin{align}
I_1(z)=\frac{(z/2)}{0!1!}+\frac{(z/2)^3}{1!2!}+\cdots~,\quad
I_2(z)=\frac{(z/2)^2}{0!2!}+\cdots~,\quad
I_3(z)=\frac{(z/2)^3}{0!3!}+\cdots~,
\end{align}
we find
\begin{align}
&\lim_{\hbar\to 0}\frac{1}{\hbar^2}
\langle\Tr\frac{\varphi^4}{4!}\,\Tr e^{u\varphi}\rangle_{\rm conn}
=\frac{\sqrt{q}^3}{0!}\biggl[\frac{1}{0!3!}3uI_3(2\sqrt{q}u)e^{ua}
+\frac{1}{1!2!}uI_1(2\sqrt{q}u)e^{ua}\biggr]\nonumber\\
&\qquad+\frac{\sqrt{q}^2a}{1!}
\biggl[\frac{1}{0!2!}2uI_2(2\sqrt{q}u)e^{ua}\biggr]
+\frac{\sqrt{q}a^2}{2!}
\biggl[\frac{1}{0!1!}uI_1(2\sqrt{q}u)e^{ua}\biggr]~,
\end{align}
by picking up the $t^4$ terms.
Therefore the one-point function of the glueball loop operator is given
as
\begin{align}
&\langle\Tr\lambda^\alpha\lambda_\alpha e^{u\varphi}\rangle
=-2\sqrt{q}^3\frac{3!}{0!3!}
\bigg[\frac{3!}{0!3!}\frac{3}{u}I_3(2\sqrt{q}u)e^{ua}
+\frac{3!}{1!2!}\frac{1}{u}I_1(2\sqrt{q}u)e^{ua}\biggr]\nonumber\\
&\qquad-2\sqrt{q}^2a\frac{3!}{1!2!}
\biggl[\frac{2!}{0!2!}\frac{2}{u}I_2(2\sqrt{q}u)e^{ua}\biggr]
-2\sqrt{q}a^2\frac{3!}{2!1!}
\biggl[\frac{1!}{0!1!}\frac{1}{u}I_1(2\sqrt{q}u)e^{ua}\biggr]~,
\end{align}
which is transformed into the glueball resolvent through the Laplace
transformation.
The Laplace transformation can be performed as
\begin{align}
\int_0^\infty due^{-(z-a)u}\frac{n}{u}I_n(2\sqrt{q}u)
=Z^n~,\label{laplace}
\end{align}
with $Z$ (and $\overline Z$ which will appear later) defined by
\begin{align}
Z=\frac{z-a-\sqrt{(z-a)^2-4q}}{2\sqrt{q}}~,\quad
\overline Z=\frac{z-a+\sqrt{(z-a)^2-4q}}{2\sqrt{q}}~.
\end{align}
Here we have used the Laplace transformation formula for the modified
Bessel function
\begin{align}
\int_0^\infty dte^{-st}I_n(\omega t)
=\frac{(s-\sqrt{s^2-\omega^2})^n}{\sqrt{s^2-\omega^2}\omega^n}~,
\end{align}
and the recursive relation of the modified Bessel function
\begin{align}
\frac{2n}{x}I_n(x)=I_{n-1}(x)-I_{n+1}(x)~.
\end{align}
Using \eqref{laplace} we easily find
\begin{align}
\langle\Tr\lambda^\alpha\lambda_\alpha\frac{1}{z-\varphi}\rangle
=-2\sqrt{q}^3\frac{3!}{0!3!}
\biggl[\frac{3!}{0!3!}Z^3+\frac{3!}{1!2!}Z\biggr]
-2\sqrt{q}^2a\frac{3!}{1!2!}\biggl[\frac{2!}{0!2!}Z^2\biggr]
-2\sqrt{q}a^2\frac{3!}{2!1!}\biggl[\frac{1!}{0!1!}Z\biggr]~.
\label{phi4result}
\end{align}

It is remarkable that the Laplace transform of the modified Bessel
function gives a linear combination of $1$ and $\sqrt{(z-a)^2-4q}$ with
the coefficient being the polynomials of $z$, which is required for the
resolvent in the maximally confining phase.
Furthermore, it behaves as $O(z^{-1})$ in the limit $z\to\infty$, as can
be seen from
\begin{align}
Z=\overline Z^{-1}~.
\label{inverse}
\end{align}
Hence, in the final step of matching our computation to the expected
result, all we have to do is to prove that the coefficient polynomial
of $1$ is $-z^3$.
This can be done without explicit calculation.
Let us first rewrite our final result \eqref{phi4result} as
\begin{align}
&-V(z)+H(z)\sqrt{(z-a)^2-4q}\nonumber\\
&\quad=-2\sqrt{q}^3\frac{3!}{0!3!}
\biggl[\frac{3!}{0!3!}Z^3+\frac{3!}{1!2!}Z\biggr]
-2\sqrt{q}^2a\frac{3!}{1!2!}\biggl[\frac{2!}{0!2!}Z^2\biggr]
-2\sqrt{q}a^2\frac{3!}{2!1!}\biggl[\frac{1!}{0!1!}Z\biggr]~.
\label{VHsqrt}
\end{align}
Then, our task is to prove $V(z)=z^3$.
To pick up $V(z)$ let us consider the ``conjugate'' of \eqref{VHsqrt}
\begin{align}
&-V(z)-H(z)\sqrt{(z-a)^2-4q}\nonumber\\
&\quad=-2\sqrt{q}^3\frac{3!}{0!3!}
\biggl[\frac{3!}{0!3!}\overline Z^3+\frac{3!}{1!2!}\overline Z\biggr]
-2\sqrt{q}^2a\frac{3!}{1!2!}\biggl[\frac{2!}{0!2!}\overline Z^2\biggr]
-2\sqrt{q}a^2\frac{3!}{2!1!}\biggl[\frac{1!}{0!1!}\overline Z\biggr]~,
\end{align}
and add up with original \eqref{VHsqrt}.
Then we can sum up the right hand side into
\begin{align}
-2V(z)=-2\sqrt{q}^3\frac{3!}{0!3!}[Z+\overline Z]^3
-2\sqrt{q}^2a\frac{3!}{1!2!}
\biggl([Z+\overline Z]^2-\frac{2!}{1!1!}\biggr)
-2\sqrt{q}a^2\frac{3!}{2!1!}[Z+\overline Z]~,
\end{align}
because of \eqref{inverse}.
We can further rewrite the result into
\begin{align}
-2V(z)=-2\sqrt{q}^3\biggl[Z+\overline Z+\frac{a}{\sqrt{q}}\biggr]^3
+2F_4~,
\label{V}
\end{align}
with $F_4$ defined as
\begin{align}
F_4=\sqrt{q}^3\biggl(\frac{3!}{1!2!}\frac{2!}{1!1!}\frac{a}{\sqrt{q}}
+\frac{3!}{3!0!}\frac{0!}{0!0!}
\biggl[\frac{a}{\sqrt{q}}\biggr]^3\biggr)~.
\end{align}
Using
\begin{align}
Z+\overline Z+\frac{a}{\sqrt{q}}=\frac{z}{\sqrt{q}}~,
\end{align}
we find \eqref{V} reduces to
\begin{align}
V(z)=z^3-F_4~.
\end{align}
To finish our proof of $V(z)=z^3$ we need to choose $a=0$ so that $F_4$
vanishes.

Note that since $V(z)$ is the first derivative of the potential (in the
matrix model terminology), we can interpret the constant part $F_4$ as
the force felt by the cut. (See Figure 2.)
Here the main contribution of $F_4$ is $F_4\sim a^3$ for large $a$ and
the subleading term depends on the width of the cut $\sqrt{q}$.
\begin{figure}[htbp]
\begin{center}
\scalebox{1.0}[1.0]{\includegraphics{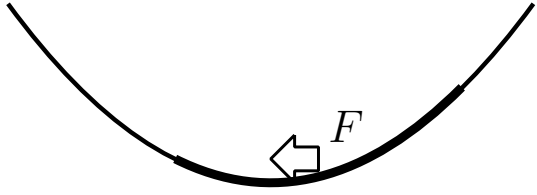}}
\setlength{\unitlength}{1mm}
\begin{picture}(20,20)
\put(-33,27){\makebox(10,10){{\Large $F$}}}
\end{picture}
\end{center}
\vspace{-25mm}
\caption{Force felt by the cut.}
\end{figure}

\subsection{Monomial superpotentials}

The above argument can be easily generalized to any monomial
superpotential.
To compute the glueball one-point function for a superpotential
\begin{align}
W(z)=\frac{z^{k+1}}{k+1}~,
\end{align}
we have to pick up the $t^{k+1}$-terms in the right hand side of
\eqref{looploop}:
\begin{align}
\langle\Tr\lambda^\alpha\lambda_\alpha e^{u\varphi}\rangle
=-2e^{ua}\sum_{l=0}^{k-1}\sqrt{q}^{k-l}a^l\,\binom{k}{l}
\sum_{m=0}^{[(k-l-1)/2]}
\binom{k-l}{m}\,\frac{k-l-2m}{u}I_{k-l-2m}(2\sqrt{q}u)~.
\end{align}
Using the Laplace transformation formula \eqref{laplace}, we find the
resolvent is given by
\begin{align}
\langle\Tr\lambda^\alpha\lambda_\alpha\frac{1}{z-\varphi}\rangle
=-2\sum_{l=0}^{k-1}\sqrt{q}^{k-l}a^l\,\binom{k}{l}
\sum_{m=0}^{[(k-l-1)/2]}\binom{k-l}{m}\,Z^{k-l-2m}~.
\label{resolventhalf}
\end{align}
Note again that the result takes the form of $-V(z)+H(z)\sqrt{z^2-4q}$.
To see the result is the expected one, we have to show $V(z)=z^k$.
This can be done by adding the ``conjugate''.
In this way, we will complete ``half'' of the binomial expansion in
\eqref{resolventhalf} into a full one except constant terms.
Finally we find $V(z)=z^k-F_{k+1}$ with the force $F_{k+1}$ given by
\begin{align}
F_{k+1}=\sum_{l=0}^{[k/2]}
\sqrt{q}^{2l}a^{k-2l}\frac{k!}{(k-2l)!(l!)^2}~.
\end{align} 
If the potential is even, the force is odd and we can always make
$F_{k+1}=0$ by setting $a=0$.
This means that in the even potential we can stabilize the cut at the
center.
However, this does not work for the odd monomial potentials.

\subsection{General superpotentials}

If the superpotential is $W_2(z)=z^2/2$, the glueball resolvent is given
as
\begin{align}
\langle\Tr\lambda^\alpha\lambda_\alpha\frac{1}{z-\varphi}\rangle
=-2\sqrt{q}\frac{1!}{0!1!}\biggl[\frac{1!}{0!1!}Z\biggr]~,
\end{align}
and the force felt by the cut is given as
\begin{align}
F_2=\sqrt{q}
\biggl(\frac{1!}{1!0!}\frac{0!}{0!0!}\frac{a}{\sqrt{q}}\biggr)~,
\end{align}
while if the potential is $W_3(z)=z^3/3$, the glueball resolvent is
\begin{align}
\langle\Tr\lambda^\alpha\lambda_\alpha\frac{1}{z-\varphi}\rangle
=-2\sqrt{q}^2\frac{2!}{0!2!}\biggl[\frac{2!}{0!2!}Z^2\biggr]
-2\sqrt{q}a\frac{2!}{1!1!}\biggl[\frac{1!}{0!1!}Z\biggr]~,
\end{align}
and the force is
\begin{align}
F_3=\sqrt{q}^2\biggl(\frac{2!}{0!2!}\frac{2!}{1!1!}
+\frac{2!}{2!0!}\frac{0!}{0!0!}
\biggl[\frac{a}{\sqrt{q}}\biggr]^2\biggr)~.
\end{align}
Hence for the potential of the linear combination
\begin{align}
V=\frac{m}{2}\varphi^2+\frac{g}{3}\varphi^3~,
\end{align}
we have to impose the force balance condition:
\begin{align}
mF_2+gF_3
=m\sqrt{q}
\biggl(\frac{1!}{1!0!}\frac{0!}{0!0!}\frac{a}{\sqrt{q}}\biggr)
+g\sqrt{q}^2\biggl(\frac{2!}{0!2!}\frac{2!}{1!1!}
+\frac{2!}{2!0!}\frac{0!}{0!0!}
\biggl[\frac{a}{\sqrt{q}}\biggr]^2\biggr)=0~.
\end{align}

As is clear from the above example the linearity holds for any
superpotential. 
Therefore, as long as we are careful with the force balance condition,
we can reproduce the results of all the superpotentials in the maximally
confining phase.

\section{Higher genus correction}

So far we have computed the classical limit of the glueball one-point
function by instanton calculus and found a complete agreement with the
matrix model result.
In the computation we have to introduce an equivariant parameter
$\hbar$ for the toric action on ${\mathbb C}^2$, which can be related to
the (local) $SO(4)$ rotations of the $\Omega$-background \cite{N}.

Note that the $\hbar$-expansion of the scalar one-point and two-point
functions computed in this paper agrees to the genus expansion of the
Gromov-Witten theory of ${\bf P}^1$ developed in \cite{OP1,OP2}.
The correspondence is described as follows:
Okounkov and Pandharipande computed all genus correlation functions of
the K\"ahler class $\omega$ and its descendents $\tau_p(\omega)$ (the
so-called stable sector) with respect to the two ramification data at
the north and the south poles of ${\bf P}^1$, which are labeled by the
partitions.
To obtain the gauge theory correlation functions we identify the
operator $\Tr\varphi^{2j}$ as the cohomology class $\tau_p(\omega)$.
Then, the $k$-instanton sector is recovered by taking both the
ramification data to be $(1^k)$.

On the other hand, the same glueball one-point function was evaluated in
\cite{DV} by relating the holomorphic quantities in supersymmetric gauge
theory to amplitudes of topological string theory \cite{BCOV} and
computing with the matrix model using open/closed string duality.
Here the loop expansion parameter of the matrix model was interpreted as
the genus expansion parameter of topological string theory, which, on
the gauge theory side, was identified with a constant graviphoton
background.
A comparison between instanton calculus and matrix model at higher
genus through topological string amplitudes was made in \cite{KMT,DST}.

Although it was pointed out \cite{T} that the $\Omega$-background and
the graviphoton background are different\footnote{The relation
between these two backgrounds was also discussed in \cite{BFFL}.},
here we would like to compute the first order correction to the
one-point function on both backgrounds and compare with each other,
because these two backgrounds look similar and share the
interpretation of genus expansion.
We shall start with computing the $\hbar$-correction of the glueball
one-point function from the instanton calculus in subsection 5.1 and
proceed to recapitulating the matrix model computation in subsection
5.2.
Comparing the result of the glueball one-point function from the
instanton calculus \eqref{gluinocorr} with the result of the one-point
function in the matrix model \eqref{loopcorr} or \eqref{exactcorr}, we
find a discrepancy.
Though we cannot find any way to relate the two results, we still expect
there to be a relation between these two genus expansions, for example,
by change of variables.

\subsection{$\hbar$-corrections of the glueball one-point function}

Let us study the $\hbar$-corrections of the glueball one-point
function.
We shall choose the simplest Gaussian superpotential here.

Since the connected two-point function is given as
\begin{align}
\langle\Tr\frac{\varphi^2}{2}\,\Tr e^{u\varphi}\rangle
=e^{ua}\hbar\sqrt{q}\,\sh(u\hbar)
I_1\biggl(2\sqrt{q}\frac{\sh(u\hbar)}{\hbar}\biggr)~,
\end{align}
we find the $\hbar$-corrections of the glueball one-point function is
\begin{align}
&\langle\Tr\lambda^\alpha\lambda_\alpha e^{u\varphi}\rangle
=-2\sqrt{q}\Bigl(1+\frac{(u\hbar)^2}{24}\Bigr)\frac{1}{u}
I_1\biggl(2\sqrt{q}u\Bigl(1+\frac{(u\hbar)^2}{24}\Bigr)\biggr)
e^{u\varphi}\nonumber\\
&\quad=-2\sqrt{q}\biggl\{\frac{1}{u}I_1(2\sqrt{q}u)e^{u\varphi}
+\frac{(u\hbar)^2}{24}2\sqrt{q}
I_0(2\sqrt{q}u)e^{u\varphi}\biggr\}~,
\end{align}
where we have used identities of the modified Bessel function
\begin{align}
I_{n-1}(x)-I_{n+1}(x)=\frac{2n}{x}I_n(x)~,\quad
I_{n-1}(x)+I_{n+1}(x)=2\frac{d}{dx}I_n(x)~.
\end{align}
Therefore the glueball resolvent is
\begin{align}
&\langle\Tr\lambda^\alpha\lambda_\alpha\frac{1}{z-\varphi}\rangle
=-2\sqrt{q}\biggl\{\frac{z-a-\sqrt{(z-a)^2-4q}}{2\sqrt{q}}
+\frac{\hbar^2}{24}2\sqrt{q}
\frac{2(z-a)^2+4q}{[(z-a)^2-4q]^{5/2}}\biggr\}~.
\label{gluinocorr}
\end{align}
Note that the following formula holds for the Laplace transformation
\begin{align}
{\cal L}\{t^nf(t)\}=(-1)^nF^{(n)}(s)~,
\end{align}
if we denote the Laplace transformation as ${\cal L}\{f(t)\}=F(s)$.

\subsection{Matrix model computation}

The computation on the matrix model side can be performed with several
methods.
The first method is due to loop equation:
The resolvent in the matrix model is determined by the loop equation.

We shall consider the matrix model with potential $W(M)$:
\begin{align}
Z=\int{\cal D}M\exp\Bigl(-N\,\Tr W(M)\Bigr)~.
\end{align}
The loop equation for the resolvent 
$R(z)=\langle N^{-1}\,\Tr(z-M)^{-1}\rangle$,
\begin{align}
\int\frac{dy}{2\pi i}\frac{W'(y)}{z-y}R(y)=R(z)^2
+\frac{1}{N^2}\frac{\delta}{\delta W(z)}R(z)~,
\end{align}
(See (13.52) in \cite{M}.) gives
\begin{align}
\int\frac{dy}{2\pi i}\frac{W'(y)}{z-y}R_1(y)=2R_0(z)R_1(z)
+\frac{\delta}{\delta W(z)}R_0(z)~,
\end{align}
at one loop.
For the gaussian model $W'(y)=\mu y$, the one-loop contribution is only
non-vanishing for $\langle\Tr\varphi^n\rangle$ with $n\ge 4$.
Hence, the resolvent for $z\to\infty$ should behave as
$R_1(z)\sim 1/z^5$.
This means there is no pole at $y=\infty$ and we can perform the $y$
integration easily by picking up the pole at $y=z$:
\begin{align}
(2R_0(z)-W'(z))R_1(z)=-\frac{\delta}{\delta W(z)}R_0(z)~.
\end{align}
The right-hand-side is a two-point loop correlator and given by
\cite{AJM}
\begin{align}
\frac{\delta}{\delta W(z)}R_0(z)=R_0(z,z)=\frac{1}{\mu(z^2-4/\mu)^2}~.
\end{align}
Therefore we find
\begin{align}
R_1(z)=\frac{1}{\mu^2(z^2-4/\mu)^{5/2}}~.
\label{loopcorr}
\end{align}

For the case of the Gaussian matrix model, an exact manipulation is
possible.
The exact solution for Gaussian matrix model
\begin{align}
\langle\frac{1}{N}\,\Tr e^{uM}\rangle=\frac{1}{Z}\int{\cal D}M
\frac{1}{N}\,\Tr e^{uM}\,\exp\Bigl(-\frac{N\mu}{2}\,\Tr M^2\Bigr)~,
\end{align}
is given by \cite{DG}
\begin{align}
\langle\frac{1}{N}\,\Tr e^{uM}\rangle
=\frac{\sqrt{\mu}}{u}I_1(2u/\sqrt{\mu})
+\frac{u^2}{12N^2\mu}I_2(2u/\sqrt{\mu})~,
\end{align}
up to genus one.
After the Laplace transformation we find
\begin{align}
\langle\frac{1}{N}\Tr\frac{1}{z-M}\rangle
=\frac{\mu}{2}(z-\sqrt{z^2-4/\mu})
+\frac{1}{N^2\mu^2}\frac{1}{(z^2-4/\mu)^{5/2}}~.
\label{exactcorr}
\end{align}

\section*{Acknowledgement}

We are grateful to F.~Cachazo, S.~Fujii, A.~Hanany, T.~Okuda, H.~Ooguri,
S.~Rey and Y.~Sumitomo for valuable discussions.
Our research is supported in part by Grant-in-Aid for Young Scientists
(B) [\# 18740143] (S.M.) and by Grant-in-Aid for Scientific Research
[\#19654007] (H.K.) from the Japan Ministry of Education, Culture,
Sports, Science and Technology.
The work of S.M. is also supported partly by Inamori Foundation and
Nishina Memorial Foundation.


\begin{thebibliography}{99}
\bibitem{SW}
N.~Seiberg and E.~Witten,
``Electric - magnetic duality, monopole condensation, and confinement in
N=2 supersymmetric Yang-Mills theory,''
Nucl.\ Phys.\  B {\bf 426}, 19 (1994)
[Erratum-ibid.\  B {\bf 430}, 485 (1994)]
[arXiv:hep-th/9407087].

\bibitem{N}
N.~A.~Nekrasov,
``Seiberg-Witten prepotential from instanton counting,''
Adv.\ Theor.\ Math.\ Phys.\  {\bf 7}, 831 (2004)
[arXiv:hep-th/0206161].

\bibitem{FMPT}
F.~Fucito, J.~F.~Morales, R.~Poghossian and A.~Tanzini,
``N = 1 superpotentials from multi-instanton calculus,''
JHEP {\bf 0601}, 031 (2006)
[arXiv:hep-th/0510173].

\bibitem{DV}
R.~Dijkgraaf and C.~Vafa,
``Matrix models, topological strings, and supersymmetric gauge
theories,''
Nucl.\ Phys.\  B {\bf 644}, 3 (2002)
[arXiv:hep-th/0206255].

\bibitem{GV}
R.~Gopakumar and C.~Vafa,
``On the gauge theory/geometry correspondence,''
Adv.\ Theor.\ Math.\ Phys.\  {\bf 3}, 1415 (1999)
[arXiv:hep-th/9811131].

\bibitem{Fer1}
F.~Ferrari,
``Microscopic quantum superpotential in N=1 gauge theories,''
JHEP {\bf 0710}, 065 (2007)
[arXiv:0707.3885 [hep-th]].

\bibitem{FKW}
F.~Ferrari, S.~Kuperstein and V.~Wens,
``Glueball operators and the microscopic approach to N=1 gauge
theories,''
JHEP {\bf 0710}, 101 (2007)
[arXiv:0708.1410 [hep-th]].

\bibitem{Fer2}
F.~Ferrari,
``Extended N=1 super Yang-Mills theory,''
arXiv:0709.0472 [hep-th].

\bibitem{FKMO}
S.~Fujii, H.~Kanno, S.~Moriyama and S.~Okada,
``Instanton calculus and chiral one-point functions in supersymmetric
gauge theories,''
[arXiv:hep-th/0702125].

\bibitem{BCOV}
M.~Bershadsky, S.~Cecotti, H.~Ooguri and C.~Vafa,
``Kodaira-Spencer theory of gravity and exact results for quantum string
amplitudes,''
Commun.\ Math.\ Phys.\  {\bf 165}, 311 (1994)
[arXiv:hep-th/9309140].

\bibitem{CDSW}
F.~Cachazo, M.~R.~Douglas, N.~Seiberg and E.~Witten,
``Chiral rings and anomalies in supersymmetric gauge theory,''
JHEP {\bf 0212}, 071 (2002)
[arXiv:hep-th/0211170].

\bibitem{NO}
N.~Nekrasov and A.~Okounkov,
``Seiberg-Witten theory and random partitions,''
arXiv: hep-th/0306238.

\bibitem{OP1}
A. Okounkov and R. Pandharipande,
``Gromov-Witten Theory, Hurwitz Theory and Completed Cycles,''
Ann.\ of Math.\  {\bf 163} (2006) 517-560,
[arXiv:math.AG/0204305].

\bibitem{OP2}
A.~Okounkov and R.~Pandharipande,
``The equivariant Gromov-Witten theory of $P^1$,''
Ann.\ of Math.\  {\bf 163} (2006) 561-605,
[arXiv:math.AG/0207233].

\bibitem{FFZ}
D.~B.~Fairlie, P.~Fletcher and C.~K.~Zachos,
``Trigonometric Structure Constants for New Infinite Algebras,''
Phys.\ Lett.\  B {\bf 218}, 203 (1989).

\bibitem{KR}
V.~Kac and A.~Radul,
``Quasifinite highest weight modules over the Lie algebra of
differential operators on the circle,''
Commun.\ Math.\ Phys.\  {\bf 157}, 429 (1993)
[arXiv:hep-th/9308153].

\bibitem{NT}
T.~Nakatsu and K.~Takasaki,
``Melting Crystal, Quantum Torus and Toda Hierarchy,''
arXiv:0710.5339 [hep-th].

\bibitem{KMT}
A.~Klemm, M.~Marino and S.~Theisen,
``Gravitational corrections in supersymmetric gauge theory and matrix
models,''
JHEP {\bf 0303}, 051 (2003)
[arXiv:hep-th/0211216].

\bibitem{DST}
R.~Dijkgraaf, A.~Sinkovics and M.~Temurhan,
``Matrix models and gravitational corrections,''
Adv.\ Theor.\ Math.\ Phys.\  {\bf 7}, 1155 (2004)
[arXiv:hep-th/0211241].

\bibitem{T}
Y.~Tachikawa, 
``Seiberg-Witten Theory and Instanton Counting'', 
Master thesis, University of Tokyo, 2003.

\bibitem{BFFL}
M.~Billo, M.~Frau, F.~Fucito and A.~Lerda,
``Instanton calculus in R-R background and the topological string,''
JHEP {\bf 0611}, 012 (2006)
[arXiv:hep-th/0606013].

\bibitem{M}
Yu.~Makeenko,
``Methods of contemporary gauge theory,''
Cambridge Univ.\ Pr.\ (2002).

\bibitem{AJM}
J.~Ambjorn, J.~Jurkiewicz and Yu.~M.~Makeenko,
``Multiloop correlators for two-dimensional quantum gravity,''
Phys.\ Lett.\  B {\bf 251}, 517 (1990).

\bibitem{DG}
N.~Drukker and D.~J.~Gross,
``An exact prediction of N = 4 SUSYM theory for string theory,''
J.\ Math.\ Phys.\  {\bf 42}, 2896 (2001)
[arXiv:hep-th/0010274].

\end{thebibliography}
\end{document}